\documentclass[twocolumn,aps,preprintnumbers,prl,amsfonts,amsmath,amssymb,superscriptaddress,floatfix]{revtex4}
\usepackage[dvips]{graphicx}
\usepackage[]{verbatim}
\allowdisplaybreaks[3]

\begin{document}

\title{Coupling of a Single Quantum Emitter to End-to-end Aligned Silver Nanowires} %Title of paper

\author{Shailesh Kumar}\email{shailesh@fysik.dtu.dk}
\author{Alexander Huck}
\affiliation{Department of Physics, Technical University of Denmark, Building 309, 2800 Kongens Lyngby, Denmark}
\author{Yuntian Chen}
\affiliation{Department of Photonics, Technical University of Denmark, Building 343, 2800 Kongens Lyngby, Denmark}
\author{Ulrik L. Andersen}
\affiliation{Department of Physics, Technical University of Denmark, Building 309, 2800 Kongens Lyngby, Denmark}

\date{\today}

\begin{abstract}
We report on the observation of coupling a single nitrogen vacancy (NV) center in a nanodiamond crystal to a propagating plasmonic mode of silver nanowires. The nanocrystal is placed either near to the apex of a single silver nanowire or in the gap between two end-to-end aligned silver nanowires. We observe an enhancement of the NV-centers' decay rate in both cases as a result of the coupling to the plasmons. The devices are nano-assembled with a scanning probe technique. Through simulations, we show that end-to-end aligned silver nanowires can be used as a controllable splitter for emission from a dipole emitter.

\end{abstract}

\maketitle

Nano-sized plasmonic structures support strongly confined electromagnetic fields at the nanometer scale. This leads to a variety of different effects such as surface enhanced Raman spectroscopy~\cite{1997Ramansinglemolecule}, the generation of higher harmonics~\cite{Kim2008} and the enhancement of fluorescence from emitters due to the Purcell effect~\cite{Purcell}. Moreover, plasmonic nanostructures can also be used as antennas, thereby tailoring the emission profile of nearby emitters~\cite{VanHulst2010}. A variety of different emitters, including molecules~\cite{2006PRLenhancementsphere, 2009Nphotonbowtie}, quantum dots~\cite{Curto2010} and nitrogen-vacancy (NV) centers in diamond~\cite{2009Schietinger} have benefitted from these plasmonics properties - enhancement and directionality of emission. However, for some applications in quantum optics, for instance single photon generation or absorption, efficient coupling between a single emitter and a single propagating mode is desirable~\cite{2007Chang}. Plasmonic waveguides support the propagation of highly confined electromagnetic fields in the direction transverse to the propagation direction, and favors an efficient coupling to nearby emitters~\cite{2010Bozhevolnyi}. This leads to an enhancement of the decay rate of the emitter, and channeling of its emission into a single propagating plasmonic mode~\cite{2006Chang, PRBChang, 2010Chen, 2007Akimov, 2009Kolesov, 2011Huck}. In particular, in Ref.~\cite{2006Chang, PRBChang} it was found that the coupling of a single emitter to the apex of a nanotip is particularly promising as in such a system a good compromise between losses and coupling is attained. 

The NV-center in a diamond has proven to be a serious candidate for quantum information processing due to its unique room-temperature properties. It emits single photons~\cite{NVnanocrystal} in the near infrared regime without blinking and its ground state forms a stable qubit with a long coherence time that can be easily initialized, manipulated and read out~\cite{1997GruberNVspinreadout,2009NVspincoherence}. However, for advancing the NV technology for e.g. long distance communication~\cite{PhysRevLett.96.070504} or distributed quantum computing~\cite{2005Lim}, the dipole moment of the NV-center must be efficiently coupled to a single propagating mode in order to direct the emitted photons in a specified direction.

In this letter, we study the controlled coupling between single NV-centers and plasmonic waveguides. Using a scanning probe technique, we nano-assemble a system consisting of a nanodiamond with a single NV-center placed in the proximity of the apex of either a single silver nanowire (SNW) or two end-to-end aligned SNWs. We demonstrate efficient coupling of the NV dipole to the plasmonic mode supported by the SNWs which is witnessed by an enhancement of the decay rate of the NV's dipole transition as well as channeling of the emitted light into the plasmonic mode.

\begin{figure}
\includegraphics[]{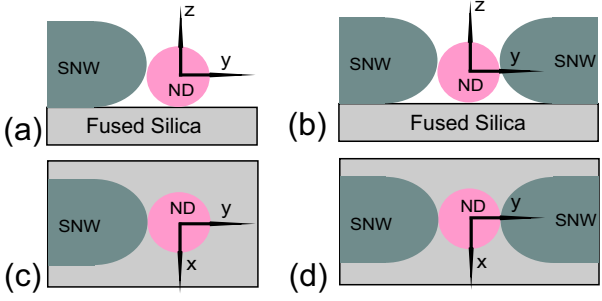}
\caption{ (Color online)\textbf{(a)} and  \textbf{(b)} Side view  schematics of a nanodiamond placed near an apex of an SNW and in the gap between two end-to-end aligned SNWs. \textbf{(c)} and \textbf{(d)} Top view schematics of the same structures as in \textbf{(a)} and  \textbf{(b)}, respectively. SNW: silver nanowire, ND: nanodiamond.}
 \label{schematics}
\end{figure}

In Fig.~\ref{schematics}, we present the schematic structure of the NV-SNW systems under consideration. The first system is an NV-center in a nanodiamond crystal coupled to the apex of a single SNW, whose side-view and top-view schematics are shown in Fig.~\ref{schematics}(a) and (c), respectively. Similarly, Fig.~\ref{schematics}(b) and (d) show the side-view and top-view schematics of the second system, which comprises an NV-center in a nanodiamond placed in between two end-to-end aligned SNWs. We note that in these figures the SNWs are cut short for presentational reasons. In the simulations and the experiments the SNWs have a length of a few $\mu$m. In the simulations, the NV-center is modeled as a dipole emitter lying inside the nanodiamond. The material surrounding the dipole emitter also affects its decay rate\cite{2011Greffet}. Therefore, we normalize the total decay rate $\Gamma$ of the coupled systems to the decay rate $\Gamma_0$ of a dipole emitter inside the nanodiamond without SNWs for different orientations and positions of the dipole.

\begin{figure}
\includegraphics[]{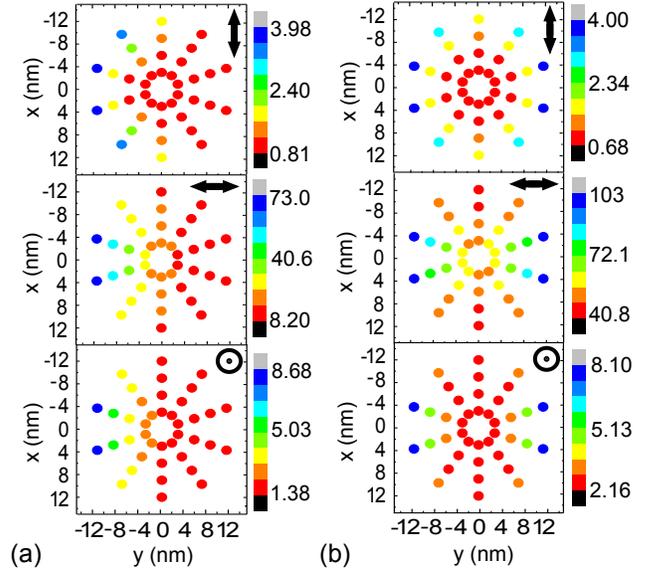}
\caption{(Color online) Simulation results showing decay rate enhancements $\Gamma/\Gamma_0$ (in color scale) as a function of position and orientation of the dipole inside the nanocrystal. $\Gamma_0$ is obtained for the same dipole emitter in a nanodiamond placed on a glass substrate without SNWs. \textbf{(a)} and \textbf{(b)} Plots of $\Gamma/\Gamma_0$ for dipole emitters lying in the xy-plane (z = 0 nm) in a nanodiamond, close to one SNW and in the gap between two end-to-end aligned SNWs, respectively. The orthogonal dipole orientations used for the simulations are indicated in various plots.}
 \label{simulations}
\end{figure}

We simulated decay rates for the systems illustrated in Fig.~\ref{schematics} using a finite element method in a commercial software (COMSOL), following a procedure developed by one of us (see Chen et al.~\cite{2010Chen}). In the simulation, the SNWs are modeled as cylinders with hemispherical ends. The radius of the SNW is taken as 25~nm and the nanodiamond is assumed to have a spherical shape with a radius of 15~nm. The distance from the center of the nanodiamond to the edge of the SNW along the y-axis is set to 15~nm. In case of end-to-end aligned SNWs, the gap between the wires equals the diameter of the nanodiamond. A vacuum wavelength of 700~nm and the corresponding refractive indices are used for the simulations~\cite{Palik}, with $n_{diamond} = 2.41$ for diamond and $n_{SiO_2} = 1.46$ for fused silica glass, respectively. The decay rate enhancements $\Gamma/\Gamma_0$ obtained from the simulations for a dipole emitter coupled to a single SNW and two end-to-end aligned SNWs are summarized in Fig.~\ref{simulations}.

Fig.~\ref{simulations}(a) shows the decay rate enhancements in the xy-plane (at z = 0 nm) for a nanodiamond lying close to an SNW apex. For the three orthogonal dipole orientations, the decay rate enhancements are plotted as a function of the dipole positions (see figure caption). For an x-oriented dipole, the decay rate can be either enhanced (up to 3.98) or suppressed (down to 0.81), depending on the dipole position inside the diamond crystal. In contrast, the emission of a y-oriented dipole, whose dipole moment is aligned with the SNW axis, the decay rates are highly enhanced (up to 73.0). For a z-oriented dipole, the decay rates are enhanced in general, but the values are smaller compared to a y-oriented dipole (between 1.38 and 8.68). Thus, it is clear that for y-oriented dipoles the highest enhancements $\Gamma/\Gamma_0$ can be obtained. 

The simulated changes of the decay rate $\Gamma/\Gamma_0$ for the end-to-end aligned system are presented in Fig.~\ref{simulations}(b).  The position dependence of the decay rate enhancements becomes symmetrical around  the xz- and yz-planes, as one might expect. Similar to the results of the single SNW system, the decay rate of an x-oriented dipole can be either enhanced (up to 4.0) or suppressed (down to 0.68), depending on the dipole location inside the nano-crystal. However, for a y-oriented dipole, the decay rate enhancements are much higher, ranging from 40.8 to 103. In general, it can be concluded from the simulations that an emitter's decay rate is enhanced in case the emitter is located close to an SNW, and it gets further enhanced when the dipole emitter is placed in the gap between two SNWs. In case of end-to-end aligned wires, the enhancement can be obtained in the entire gap region for the right orientation of the dipole. In the following, we will verify this pattern of decay rate enhancement experimentally.

The experiment is carried out with a home built confocal fluorescence microscope in combination with an atomic force microscope (AFM). We use an oil immersion objective (NA = 1.4) to focus the pump beam with a wavelength of 532~nm (cw or pulsed with 4.6~ps pulse width and 5.05~MHz pulse rate) onto our sample. Fluorescence light, collected with the same objective, is split via a symmetric beam splitter into two arms. One of the beam splitter output ports is imaged onto a pinhole for spatial selection and recorded with an avalanche photodiode (APD). The other beam splitter output port contains in addition galvanometric mirrors which can be used to scan the image plane of the objective while keeping the excitation spot fixed. Using optical filters in front of the APDs, we select the fluorescence light between 647~nm and 785~nm, thereby discarding the green excitation light. A spectrometer can be used for spectral detection instead of one of the APDs. The AFM cantilever tip is coaxially aligned with the confocal microscope and is used to image and manipulate the SNWs and nanodiamonds.

\begin{figure*}[htbp]
\includegraphics[]{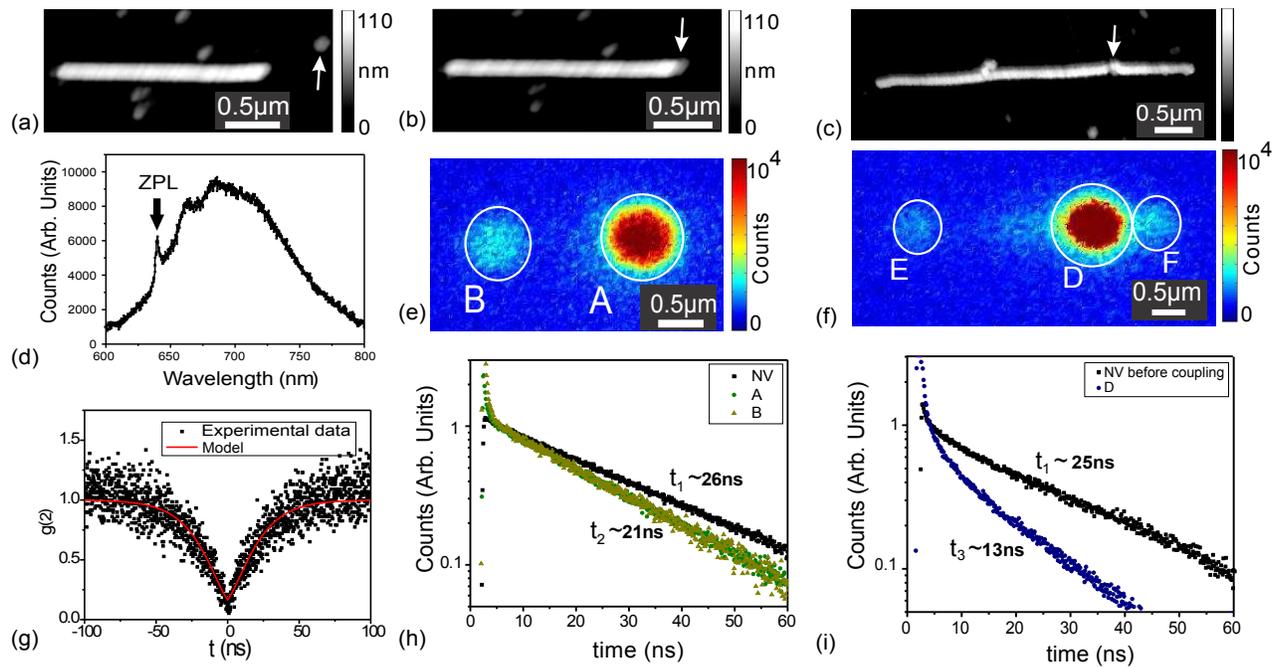}
\caption{(Color online) \textbf{(a)-(c)} AFM images of silver nanowires and a single NV-center contained in a nanodiamond which is indicated by the arrows in respective figures.  \textbf{(d)} Emission spectrum from the nanodiamond indicated in (a). ZPL: zero phonon line. \textbf{(e), (f)} Fluorescence images of the nanowire-nanodiamond systems shown in (b) and (c) respectively, where the excitation point is the nanodiamond containing a single NV-center. \textbf{(g)} Autocorrelation measurement data and a model fit to the data obtained  for emission from the nanodiamond indicated in (a).  \textbf{(h)} and \textbf{(i)} Graph showing the measurement data for the change in lifetime when the nanodiamond containing NV-center is moved near to silver wires as shown in (b) and (c), respectively. (c), (f) and (i) correspond to system 2 in table I, whereas the other figures correspond to system 1 in table I.}
\label{Experiment}
\end{figure*}

The samples were made by spin coating a solution containing nanodiamonds (Microdiamant MSY 0-0.1), followed by a solution of SNWs (made with a polyol reduction process of silver nitrate~\cite{Agwirefabrication}) onto a plasma cleaned fused silica substrate. We identified nanodiamonds containing a single NV-center lying close to the SNWs by matching the location of the fluorescence from the cantilever tip and fluorescence from the nanodiamond. In Fig.~\ref{Experiment}(a) we present an AFM image of an SNW and the nanodiamond containing a single NV-center indicated by an arrow. The fluorescence spectrum and measured second-order correlation function $g^{(2)}(\tau)$ for the nanodiamond indicated in Fig.~\ref{Experiment}(a) are presented in Fig.~\ref{Experiment}(d) and (g), respectively. $g^{(2)}(\tau)$ has been modeled by a 3-level system~\cite{Threelevelmodel} and the value $g^{(2)}(\tau=0) <0.5$ indicates the emission of single photons.

\begin{table*}
  \caption{Decay rate enhancements due to coupling to one SNW and two end-to-end aligned SNWs, normalized to the decay rate of the NV-centers without SNWs.}
  \label{tbl:twoNW}
  \begin{tabular}{ c  c  c  c  c  c }
    \hline
    System  & SNW1 &  SNW2& ND  & $\Gamma/\Gamma_0$ & $\Gamma/\Gamma_0$ \\
   number & diameter (nm)  & diameter(nm) & height (nm) &  (1 wire)  &  (2 wires)\\
    \hline
    1 & 72 & 72  & 49 & 1.23 & 1.54 \\
    2 & 80 & 80  & 59 & 1.28 & 1.94 \\
    3 & 95 & 115 & 48 & 1.61 & 2.28 \\
    \hline
  \end{tabular}
\end{table*}

Using the AFM in contact mode, the nanodiamond was then pushed across the sample close to one end of an SNW. Fig.~\ref{Experiment}(b) shows an AFM image of such a nanodiamond-SNW system. A fluorescence image of this system recorded while keeping the excitation spot fixed at the NV-center position is shown in Fig.~\ref{Experiment}(e). One can clearly identify two spots, which are labeled as A and B in the figure. Spot A originates dominantly from the NV-center emitting into free space modes whereas spot B  results from the coupling of the NV-center fluorescence into the plasmonic mode of the silver nanowire, plasmon propagation to the distal wire end and the subsequent scattering of the plasmon to the far field. The lifetime measurements of this NV-center before coupling (indicated in Fig.~\ref{Experiment}(a)) and after coupling (taken from spots A and B in Fig.~\ref{Experiment}(e)) are summarized in Fig.~\ref{Experiment}(h). It can be seen that the coupled system is characterized by two lifetime slopes (both for A and B). The very fast decay ($\tau < 1 ns$) corresponds to fluorescence from the SNW while the slow decay curve corresponds to fluorescence from the NV-center. By taking the ratio of the lifetimes of NV-contributions, we observe an increase of the emission rate for this single wire case by a factor of approximately 1.23. We note that the slow decay curve obtained from both spots, A and B, are the same (within the small uncertainty of our measurement), indicating that the coupling of the NV center to the propagating mode of the SNW is significant.

After coupling the NV-center to the end of a single SNW, we placed a second SNW close to the nanodiamond so that the two SNWs are aligned along a straight line with a small gap left between the ends. The second SNW is either cut from the first nanowire or it is a different nanowire. To cut an SNW, we moved the AFM cantilever tip across the SNW with a contact force of around~$1~\mu$N at a velocity of a few $\mu$m/s. The SNWs were moved slowly by pushing them at 5-10 positions with a contact force of approximately~$0.1~\mu$N with the AFM cantilever. Fig.~\ref{Experiment}(c) shows the AFM image of two wires aligned along a straight line with a nanodiamond placed in the gap between the two wires (System 2 in table I). In the fluorescence image shown in Fig.~\ref{Experiment}(f), one can observe the fluorescence from three spots; D, E and F, where D is the spot corresponding to radiative emission from the NV-center, and E and F are two distal ends of the two SNWs. We also measure the lifetime after the end-to-end aligned system is prepared, which shows a reduced lifetime for the NV-center as presented in Fig.~\ref{Experiment}(i). The reduction in lifetime as well as the emission from two distal ends confirms the coupling of NV-center emission to the end-to-end aligned nanowire system.

In Table~\ref{tbl:twoNW}, we summarize the decay rate enhancements observed for different systems labeled by 1, 2 and 3. In system 1 and 2, the radii of the two wires are the same, while in system 3, the second wire has a slightly larger diameter. In all realizations, we observed a decay rate enhancement of the NV-center with respect to the bare glass-air interface as well as the single SNW-NV-center system. Not all realizations resulted in clear emission from the two distal ends, even though we observed an increase in decay rate in each realization. This could be either due to assymetric coupling to the two wires or small coupling and subsequent propagation losses in the plasmonic wires.

\begin{figure}[htbp]
\includegraphics[]{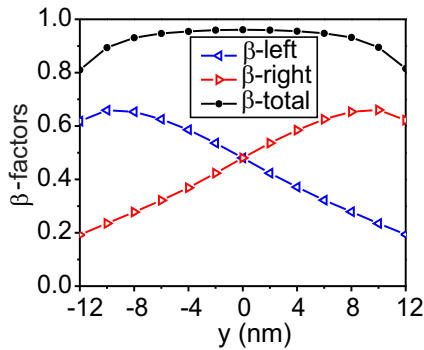}
\caption{(Color online) Spontaneous emission $ \mathbf{\beta} $ - factor for coupling to the two SNWs which are aligned end-to-end. Blue and red curves shows $\beta$ - factors for coupling to the left ($\beta_{left}$) and the right ($\beta_{right}$) SNWs, respectively. Black curve shows the total $\beta$ - factor ($\beta_{left} + \beta_{right}$).  \label{beam_splitter}}
\end{figure}

Up to this point, we mainly focused our attention on changes of the NV-centers decay rate. In the context of realistic applications, an additional parameter of importance for such an NV-center plasmon coupled system is the spontaneous emission $\beta$-factor. The $\beta$-factor is defined as the ratio of emission into a particular mode relative to the total emission including all modes. In Fig.~\ref{beam_splitter}, we present theoretical predictions on the spontaneous emission $\beta$-factor as a function of NV-center position inside the diamond; $\beta_{left}$ ($\beta_{right}$) quantifies the emission into the left (right) SNW and $\beta_{total}$ = $\beta_{left}$ + $\beta_{right}$. The SNW diameter in this case is 25~nm and the gap between the two SNWs is 34~nm. Clearly, emission into a particular SNW can be controlled by shifting the position of the emitter. This result is in contrast to the coupling of an emitter to a single SNW where the emitter is placed on the side of the plasmonic waveguide~\cite{2006Chang, PRBChang}, in which case the emission is always symmetrical in two opposite directions. 

In conclusion, we have shown theoretically that the coupling of a dipole emitter to the apex of a single SNW as well as two end-to-end aligned SNWs can be high. We found that the end-to-end wire system is particularly promising, and that the coupling strength to propagating plasmonic modes critically depends on the orientation and the position of the dipole with respect to the wires. The two configurations were investigated experimentally, and we observed an enhancement in both cases but the end-to-end wire system yielding the largest decay rate enhancement. The two systems were assembled consecutively using the same constituents which enabled a direct comparison of the decay rates. We also verified that the propagating plasmonic modes were excited by the NV-center. Finally, we note that the reported experimental results on coupling an NV-center to SNWs can be substantially enhanced by improving the structuring of the wire ends and by placing the NV-centers at optimized positions.

\begin{acknowledgments}

The authors acknowledge financial support by the Villum Kann Rasmussen foundation, the Carlsberg foundation and the Danish Research Council for Technology and Production Sciences (grant number 10093787).
\end{acknowledgments}

\end{document}